\documentclass[10pt]{amsart}
\usepackage{amssymb}
\usepackage{enumerate}
\usepackage[dvips]{graphicx}
\DeclareGraphicsExtensions{.eps,.art,.ART,.ps}
\newcommand{\bbR}{\mathbb{R}}      % real numbers %
\newcommand{\tr}{\operatorname{tr}}
\newcommand{\cotan}{\operatorname{cotan}}
\newcommand{\cotanh}{\operatorname{cotanh}}
\newtheorem{Theo}{Theorem}[section]
\newtheorem{Prop}[Theo]{Proposition}

\theoremstyle{definition}
\newtheorem{Def}[Theo]{Definition}
\theoremstyle{remark}

\begin{document}
\begin{abstract}
We find a class of warp drive spacetimes possessing Newtonian limits, which we then determine. The same method is used to compute Newtonian limits of the Schwarzschild solution and spatially flat Friedmann-Robertson-Walker cosmological models.
\end{abstract}
%
%%%%%%%%%%%%%%%%%%%%%%%%%%%% Definition of title page %%%%%%%%%%%%%%%%%%%%%%%%%%%
%
\title{Newtonian limits of warp drive spacetimes}
\author{Jos\'{e} Nat\'{a}rio}
\address{Department of Mathematics, Instituto Superior T\'{e}cnico, Portugal}
\email{jnatar@math.ist.utl.pt}
%\subjclass[2000]{Primary 53078}
\thanks{This work was partially supported by FCT/POCTI/FEDER}
\maketitle
%
%
%%%%%%%%%%%%%%%%%%%%%%%%%%%%%%%%%%% Introduction %%%%%%%%%%%%%%%%%%%%%%%%%%%%%%%%%%
%
\section*{Introduction}
Newtonian limits of relativistic spacetimes are useful tools for understanding and interpreting these (see \cite{E91}, \cite{E97} and references therein). The warp drive spacetime, first introduced in \cite{A94}, is an interesting example of a relativistic spacetime, sheding light on the nature and limits of General Relativity (see for instance \cite{LV04} and references therein), but its interpretation as a gravitational field is somewhat unclear. It would therefore be desirable to obtain a Newtonian limit of this spacetime; this is the problem we address.

This paper is divided into five sections. In the first section, we quickly review the initial value formulation of General Relativity, including formulae which we will need in what follows. In the second section, we define Newtonian spacetime and its associated potential function, and show that there exist remarkable similarities between these spacetimes and their counterparts in Newtonian theory. These similarities are even more impressive in the important case in which the source is a dust (with cosmological constant), which we examine in the third section. In the fourth section, we show that we can always obtain a Newtonian limit for a Newtonian spacetime corresponding to the {\em same} potential function; we do this in the fifth section for warp drive Newtonian spacetimes, the Schwarzschild solution and spatially flat Friedmann-Robertson-Walker cosmological models.

We shall use the notation and sign conventions in \cite{W84}, where latin indices represent abstract indices; we will however reserve the latin indices $i,j,\ldots$ for {\em numerical} indices referring to the space manifold (i.e., $i,j,\ldots = 1,2,3$). Whenever a tensor field is described by its components, these will always be relative to the Eulerian observers' frame defined in the first section. As is usual, we will not worry about the vertical position of indices when using Cartesian coordinates.
%
%
%%%%%%%%%%%%%%%%%%%%%%%%%%%%%%%%%%% Section 1 %%%%%%%%%%%%%%%%%%%%%%%%%%%%%%%%%%
%
\section{Synchronized observers}
Let $(M,g)$ be a $4$-dimensional Lorentzian manifold, $\Sigma \subset M$ a spacelike hypersurface and $n^a$ the unit normal vector field to this surface. Taking $n^a$ as the initial condition for a (timelike) geodesic, we can construct a family of free-fall observers (which we shall call {\em Eulerian observers}). The proper time of each observer along the geodesic (starting from $\Sigma$) defines a time function $t$ in an open neighbourhood of $\Sigma$. Shrinking $\Sigma$ if necessary, we can assume that $t$ foliates some open set $U$ by spacelike surfaces $\Sigma_{t_0}=t^{-1}(t_0)$, with $\Sigma_0=\Sigma$, diffeomorphic by the flow of $n^a$. As is well known, the hypersurfaces $\Sigma_{t_0}$ are orthogonal to $n^a$, and hence integrate the distribution of hyperplanes defined by the local notions of simultaneity of the Eulerian observers; we then say that the Eulerian observers are {\em synchronized} -- see \cite{LL97}).

The integral lines of $n^a$ determine a natural projection $\pi:U \to \Sigma$. Hence local coordinates $\xi^i$ on an open subset $V\subset \Sigma$ yield local coordinates $(t,\xi^i)$ on $\pi^{-1}(V)\subset U$. In these coordinates,
\[
g = - dt\otimes dt + \gamma_{ij}(t)\, d\xi^i\otimes d\xi^j.
\]
Notice that
\[
\gamma(t) = \gamma_{ij}(t) \, d\xi^i\otimes d\xi^j
\]
can be thought of as a (time-dependent) Riemannian metric on $\Sigma$. We shall call $(\Sigma, \gamma(t))$ the {\em space manifold at time} $t$.

It is a simple matter to compute the components of the Riemann curvature tensor $R_{abc}^{\,\,\,\,\,\,\,\, d}$ and of the Ricci tensor $R_{ab}$ of $(M,g)$ on these coordinates: they are
\begin{align}
\label{Riemann1} & R_{0i0}^{\,\,\,\,\,\,\,\, j} = - \frac{\partial}{\partial t} K^{j}_{\,\, i} - K_{il} K^{lj}; \\ 
\nonumber & R_{ij0}^{\,\,\,\,\,\,\,\, l} = - D_i K^l_{\,\, j} + D_j K^{l}_{\,\, i}; \\ 
\nonumber & R_{ijl}^{\,\,\,\,\,\,\,\, m} = \hat{R}_{ijl}^{\,\,\,\,\,\,\,\, m} + K_{il} K^{m}_{\,\,\,\, j} - K_{jl} K^{m}_{\,\,\,\, i};
\end{align}
and
\begin{align*}
& R_{00} = - \frac{\partial}{\partial t} K^{i}_{\,\, i} - K_{ij} K^{ij}; \\
& R_{0i} = - D_i K^j_{\,\, j} + D_j K^{j}_{\,\, i}; \\
& R_{ij} = \hat{R}_{ij} + \frac{\partial}{\partial t} K_{ij} - 2 K_{il} K^{l}_{\,\, j} + K^{l}_{\,\, l} K_{ij},
\end{align*}
where $\hat{R}_{ijl}^{\,\,\,\,\,\,\,\, m}=\hat{R}_{ijl}^{\,\,\,\,\,\,\,\, m}(t)$ and $\hat{R}_{ij}=\hat{R}_{ij}(t)$ are the components of the Riemann and Ricci tensors of the Levi-Civita connection $D = D(t)$ of $\gamma(t)$, and
\[
K_{ij}(t) = \frac12 \frac{\partial}{\partial t} \gamma_{ij}(t).
\]
In these calculations, one must bear in mind that raising and lowering indices with the space manifold metric $\gamma(t)$ {\em does not} commute with $\frac{\partial}{\partial t}$, since
\[
\frac{\partial}{\partial t} \gamma_{ij} = 2K_{ij}, \,\,\,\,\,\,\,\, \frac{\partial}{\partial t} \gamma^{ij} = -2K^{ij}.
\]
For instance, equation (\ref{Riemann1}) can be rewritten as
\begin{align}
\label{Riemann1'} & R_{0i0}^{\,\,\,\,\,\,\,\, j} = - \frac{\partial}{\partial t} \left(\gamma^{jl}K_{li}\right) - K_{il} K^{lj} = - \gamma^{jl}\frac{\partial}{\partial t}K_{li} + K_{il} K^{lj}.
\end{align}

Again,
\[
K(t) = K_{ij}(t) \, d\xi^i\otimes d\xi^j
\]
can be thought of as a (time-dependent) symmetric tensor on $\Sigma$. On the other hand, both $\gamma$ and $K$ can be thought of as tensor fields on $U$, by considering $\gamma(t)$, $K(t)$ to be tensor fields on each submanifold $\Sigma_t \subset U$ and then extending them trivially to the direction given by $n^a$:
\begin{align*}
& \gamma_{ab} = g_{ab} + n_a n_b; \\
& K_{ab} = \frac12 \pounds_n \gamma_{ab} = \frac12 \pounds_n g_{ab} = \frac12 \left( \nabla_a n_b + \nabla_b n_a \right).
\end{align*}
Notice that $K_{ab}$ yields the extrinsic curvature of the family $\Sigma_t$.
%
%
%%%%%%%%%%%%%%%%%%%%%%%%%%%%%%%%%%%%%%% Section 2 %%%%%%%%%%%%%%%%%%%%%%%%%%%%
%
\section{Newtonian spacetimes}
If all the hypersurfaces $\Sigma_t$ are flat, then one can easily find coordinates $(t,x^i)$ in $U$ such that $(x^i)$ are Cartesian coordinates\footnote{Using for instance the exponential map on these hypersurfaces. The coordinates $x^i$ are sometimes called {\em Eulerian coordinates}, whereas $\xi^i$ are known as the {\em Lagrangian coordinates}. It would be perhaps more reasonable to call the Eulerian observers 'Lagrangian observers', as it is their Lagrangian coordinates which remain constant; nevertheless, we shall keep with the common General Relativity usage.} on $\Sigma_t$: 
\begin{equation} \label{linelement}
g = - dt \otimes dt + \delta_{ij} (dx^i - v^i dt) \otimes (dx^j - v^j dt).
\end{equation}
Notice that on these coordinates one has
\[
n^a = \frac{\partial}{\partial t} + v^i \frac{\partial}{\partial x^i};
\]
therefore, ${\bf v}=v^i \frac{\partial}{\partial x^i}$ can be intrepreted as the $3$-velocity of the Eulerian observers on a fixed Euclidean background. Since the coordinates $x^i$ are are only defined up to a (time-dependent) Euclidean isometry, ${\bf v}$ is only defined up to the addition of an arbitrary (time-dependent) Euclidean Killing vector field $\boldsymbol{\omega}(t) \times {\bf x} + {\bf x}_0(t)$, which is, of course, irrelevant when computing
\[
K = \frac12 \pounds_n \gamma = \frac12 (D_i v_j + D_j v_i) dx^i \otimes dx^j.
\]
If ${\bf v}$ can be chosen to be a gradient\footnote{Note that this will fix ${\bf v}$ up to a (time-dependent) additive constant ${\bf x}_0(t)$.}, ${\bf v} = \nabla \psi$, then the Eulerian observers will follow Newtonian free-falling motions on the potential
\[
\phi = - \frac{\partial \psi}{\partial t} - \frac12 {\bf v}^2,
\]
as in this case $D_i v^j = D_j v^i$ and hence
\begin{equation} \label{motion}
-D_i \phi = \frac{\partial}{\partial t} \left(D_i\psi\right) + v^j D_i v^j = \frac{\partial v^i}{\partial t} + v^j D_j v^i
\Leftrightarrow \frac{\partial {\bf v}}{\partial t} + ({\bf v} \cdot \nabla) {\bf v} = - \nabla \phi.
\end{equation}

\begin{Def}
We say that a metric of the form (\ref{linelement}) with ${\bf v} = \nabla \psi$ defines a {\em Newtonian spacetime} with potential $\phi = - \frac{\partial \psi}{\partial t} - \frac12 {\bf v}^2$.
\end{Def}

\begin{Prop}
The components of the Riemann and Ricci tensor for a Newtonian spacetime are
\begin{align}
\label{Riemann1N} & R_{0i0}^{\,\,\,\,\,\,\,\, j} = D^j D_i \phi; \\ 
\label{Riemann2N} & R_{ij0}^{\,\,\,\,\,\,\,\, l} = 0; \\ 
\label{Riemann3N} & R_{ijl}^{\,\,\,\,\,\,\,\, m} = K_{il} K^{m}_{\,\,\,\, j} - K_{jl} K^{m}_{\,\,\,\, i};
\end{align}
and 
\begin{align}
\nonumber & R_{00} = \nabla^2 \phi; \\
\nonumber & R_{0i} = 0; \\
\label{Ricci3N} & R_{ij} = -D_iD_j \phi - K_{il} K^{l}_{\,\, j} + K^{l}_{\,\, l} K_{ij}.
\end{align}
\end{Prop}

\begin{proof}
For a Newtonian spacetime $K_{ij} = D_i v^j$. Equation (\ref{motion}) then yields
\[
-D_iD_j \phi = \frac{\partial}{\partial t} K_{ij} + K_{il} K^{l}_{\,\, j} + v^l D_l K_{ij},
\]
from which
\begin{align}
\nonumber & \left(\pounds_n K\right)_{ij} = \frac{\partial}{\partial t} K_{ij} + v^l D_l K_{ij} + 2K_{il} K^{l}_{\,\, j} = -D_iD_j \phi + K_{il} K^{l}_{\,\, j};\\
\label{trace} & \pounds_n\left(K^{i}_{\,\, i}\right) = \frac{\partial}{\partial t} K^{i}_{\,\, i} + v^l D_l K^{i}_{\,\, i} = -\nabla^2 \phi - K_{ij} K^{ij}.
\end{align}
Using (\ref{Riemann1'}), the result follows.
\end{proof}

Newtonian spacetimes share a number of common features with their Newtonian counterparts:

\begin{Prop} \label{Poisson}
The potential $\phi$ satisfies the Poisson equation
\[
\nabla^2 \phi = 4\pi(T_{00} +  \gamma^{ij} T_{ij}) - \Lambda.
\]
here $T_{ab}$ is the energy-momentum tensor and $\Lambda$ is the cosmological constant.
\end{Prop}

\begin{proof}
This is just the ($00$) component of Einstein's equation
\begin{equation} \label{Einstein}
G_{ab} + \Lambda g_{ab} = 8\pi T_{ab} \Leftrightarrow R_{ab} = 8\pi \left( T_{ab} - \frac12 T^c_{\,\,c} \, g_{ab} \right) + \Lambda g_{ab}.
\end{equation}
\end{proof}

\begin{Prop} \label{potential}
Free-falling particles on a Newtonian spacetime follow, to first order on their velocities with respect to the Eulerian observers, Newtonian free-falling motions with respect to the potential $\phi$. 
\end{Prop}

\begin{proof}
Using $t$ as the parameter, the Lagrangian for free-falling motions on a Newtonian spacetime is
\[
L = \sqrt{1-\delta_{ij}\left( \frac{dx^i}{dt} - v^i \right)\left( \frac{dx^j}{dt} - v^j \right)}.
\]
To second order on the velocity with respect to the Eulerian observers this is
\[
L = - \frac12\delta_{ij}\left( \frac{dx^i}{dt} - v^i \right)\left( \frac{dx^j}{dt} - v^j \right).
\]
The Lagrangean for a Newtonian free-falling motion with respect to the potential $\phi$ is
\[
L' = \frac12 \delta_{ij} \frac{dx^i}{dt}\frac{dx^j}{dt} - \phi.
\]
Using the definition of $\phi$, we see that
\[
L + L' = v^i \frac{dx^i}{dt} + \frac{\partial \psi}{\partial t} = D_i \psi \frac{dx^i}{dt} + \frac{\partial \psi}{\partial t} = \frac{d \psi}{dt},
\]
and hence $L$ and $L'$ determine the same motion equations.
\end{proof}

\begin{Prop}
If a Newtonian spacetime is flat then the potential is an affine function.
\end{Prop}

\begin{proof}
This is immediate from equation (\ref{Riemann1N}).
\end{proof}

Surprisingly, the converse is not true. For instance, taking ${\bf v}=\frac1t {\bf x}$ one obtains $\phi=0$ and $K=\frac1t I$, and hence
\begin{align} \label{example}
\left( R_{\mu \nu} \right) =
\left(
\begin{matrix}
0 & 0\\
0 & \frac2{t^2} I
\end{matrix}
\right)
\Rightarrow
\left( G_{\mu \nu} \right) =
\left(
\begin{matrix}
\frac3{t^2} & 0\\
0 & -\frac1{t^2} I
\end{matrix}
\right).
\end{align}
This Newtonian spacetime corresponds to a cosmological solution with a negative pressure perfect fluid which nevertheless satisfies (just) both the strong and the weak energy conditions.
%
%
%%%%%%%%%%%%%%%%%%%%%%%%%%%%%%%%%%%%%%% Section 3 %%%%%%%%%%%%%%%%%%%%%%%%%%%%
%
\section{Dust Newtonian spacetimes}
To get a closer analogue to the Newtonian situation, we consider the case when the source of the gravitational field described by a Newtonian spacetime is a dust comoving with the Eulerian observers,
\[
T_{ab} = \rho \, n_a n_b
\]
(this includes vacuum solutions; see also \cite{T66}). In this case, the Poisson equation in Proposition \ref{Poisson} is simply
\begin{align} \label{PoissonD}
\nabla^2 \phi = 4\pi\rho - \Lambda,
\end{align}
which is exactly the Newtonian Poisson equation with mass density $\rho$ and cosmological constant $\Lambda$. We also have

\begin{Prop}
For Newtonian dust fields with rest density $\rho$, the continuity equation
\[
\frac{\partial \rho}{\partial t} + \nabla \cdot (\rho \, {\bf v}) = 0
\]
holds.
\end{Prop}

\begin{proof}
Since $n^a$ is geodesic, one has
\begin{align} \label{continuity}
\nabla^b T_{ab} = 0 \Leftrightarrow n^b \nabla_b \rho + \rho \, \nabla_b n^b = 0 \Leftrightarrow \pounds_n \rho + \rho \tr K = 0,
\end{align}
which can be recast as the usual continuity equation:
\[
\frac{\partial \rho}{\partial t} + ({\bf v} \cdot \nabla) \rho + \rho \, \nabla \cdot {\bf v} = 0 \Leftrightarrow \frac{\partial \rho}{\partial t} + \nabla \cdot (\rho \, {\bf v}) = 0.
\]
\end{proof}

Dust fields Newtonian spacetimes behave satisfactorily in another respect:

\begin{Prop}
If the potential associated to a Newtonian dust field is an affine function then the solution is flat.
\end{Prop}

\begin{proof}
Suppose that $H\phi = 0$. Then $\nabla^2 \phi = 0$, and hence from Poisson's equation (\ref{PoissonD}) one has $4\pi\rho=\Lambda$. 

If $\Lambda\neq 0$, one has from (\ref{continuity}) that $\tr K = 0$. Equation (\ref{trace}) then yields $\tr K^2 = 0 \Rightarrow K = 0$, and it is clear from (\ref{Riemann1N}), (\ref{Riemann2N}) and (\ref{Riemann3N}) that the solution is flat.

If $\Lambda=0$, one has $\rho=0$ and Einstein's equation (\ref{Einstein}) yields $R_{\mu \nu} = 0$. It follows from (\ref{Ricci3N}) that $K^2 - (\tr K) K = 0$. From (\ref{Riemann3N}) we see that $R_{ijl}^{\,\,\,\,\,\,\,\, j} = 0$; since $R_{ijl}^{\,\,\,\,\,\,\,\, m}$ has the same symmetries of the Riemann tensor of a $3$-dimensional Riemannian manifold, we conclude that $R_{ijl}^{\,\,\,\,\,\,\,\, m} = 0$. Finally, (\ref{Riemann1N}) and (\ref{Riemann2N}) yield $R_{0i0}^{\,\,\,\,\,\,\,\, j} = R_{ij0}^{\,\,\,\,\,\,\,\, l} = 0$.
\end{proof}
%
%
%%%%%%%%%%%%%%%%%%%%%%%%%%%%%%%%%%%%%%% Section 4 %%%%%%%%%%%%%%%%%%%%%%%%%%%%
%
\section{Newtonian limits of Newtonian spacetimes}
To construct a Newtonian limit of a Newtonian spacetime we consider the family
\begin{align}
\label{family} g(\lambda) = - \frac1\lambda dt \otimes dt + \delta_{ij} (dx^i - v^i dt) \otimes (dx^j - v^j dt)
\end{align}
(see \cite{E91}, \cite{E97}). 

\begin{Prop}
Family (\ref{family}) has a Newtonian limit corresponding to a gravitational potential $\phi$. (In other words, Newtonian spacetimes always have Newtonian limits with the {\em same} potentials).
\end{Prop}

\begin{proof} This is immediate from Proposition \ref{potential}, which shows that the leading terms of the Christoffel symbols for the Levi-Civita connection of a Newtonian spacetime are the Christoffel symbols for the Cartan connection corresponding to a gravitonal potential $\phi$. As confirmation, we compute the components of the Riemann curvature tensor for this family:
\begin{align*}
& R_{0i0}^{\,\,\,\,\,\,\,\, j} = - D^j D_i \phi; \\ 
& R_{ij0}^{\,\,\,\,\,\,\,\, l} = 0; \\ 
& R_{ijl}^{\,\,\,\,\,\,\,\, m} =  \lambda K_{il} K^{m}_{\,\,\,\, j} - \lambda K_{jl} K^{m}_{\,\,\,\, i};
\end{align*}
As $\lambda \to 0$ these clearly converge pointwise to the components of the Riemann tensor of the Cartan connection corresponding to a gravitational potential $\phi$.
\end{proof}

Note that in general both the Newtonian spacetimes (\ref{family}) and their Newtonian limits are unphysical, in the sense that they do not come with a corresponding matter model generating the field (this must be so to accomodate the warp drive). In the particular case of a dust, however, they do yield a genuine physical Newtonian limit in the sense of \cite{E91}, \cite{E97}.
% 
%
%%%%%%%%%%%%%%%%%%%%%%%%%%%%%%%%%%%%%%% Section 5 %%%%%%%%%%%%%%%%%%%%%%%%%%%%
%
\section{Examples}
% 
%%%%%%%%%%%%%%%%%%%%%%%%%%%%%%%%%%%%%%% Subsection 5.1 %%%%%%%%%%%%%%%%%%%%%%%%%%%%
%
\subsection{Newtonian warp drive}
A {\em warp drive spacetime} containing a warp bubble moving with speed $u(t)$ is given by a metric of the form (\ref{linelement}) with
\[
{\bf v}=
\begin{cases}
{\bf 0} & \text{ if } r < r_0 - \varepsilon \\
-u(t)\frac{\partial}{\partial x^1} & \text{ if } r > r_0 + \varepsilon
\end{cases}
\]
where $r^2 = \left(x^1\right)^2+\left(x^2\right)^2+\left(x^3\right)^2$ and $r_0 > \varepsilon > 0$ (this is in the frame where the bubble is at rest; see \cite{N02a} for deatils). If ${\bf v}=\nabla \psi$, this will also be a Newtonian spacetime. Choosing
\[
\psi = - u(t) f(r) x^1,
\]
where $f:\bbR \to [0,1]$ is a smooth function satisfying
\[
\begin{cases}
f(r) = 0 \text{ if } r < r_0 - \varepsilon \\
f(r) = 1 \text{ if } r > r_0 + \varepsilon
\end{cases}
\]
it is easily seen that
\[
{\bf v} = \nabla \psi = -uf\frac{\partial}{\partial x^1} - uf'x^1 \nabla r
\]
satisfies the above requirements. The corresponding potential is
\[
\phi = -\frac{\partial \psi}{\partial t} - \frac12 {\bf v}^2 = \dot{u} f x^1 - \frac{u^2}2 \left( f^2 + \left( f' x^1 \right)^2 + 2 f f' \frac{\left( x^1 \right)^2}{r} \right).
\]

This potential contains two terms: an acceleration term, which is only present when the velocity $u(t)$ is changing, and a steady term, which must be present whenever the warp drive is on ($u(t) \neq 0$).

One has $\phi = 0$ for $r < r_0 - \varepsilon$ and $\phi = \dot{u} x^1 - \frac12 u^2$ for $r > r_0 + \varepsilon$. Thus outside the bubble wall region one has vacuum and, if the bubble is not accelerating, vanishing gravitational fields. An uniform gravitational field of magnitude $\dot{u}$ is responsible for accelerating the bubble (thus guaranteeing zero spacetime curvature outside the bubble).

By ordinary potential theory, one knows that $\nabla^2\phi$ cannot be identically zero, and that moreover
\[
\int_{\bbR^3} \frac{\nabla^2\phi({\bf x})}{\|{\bf x}-{\bf x}_0\|} d^3x = 0
\]
for all ${\bf x}_0$ satisfying $\|{\bf x}_0\| > r_0 + \varepsilon$. Therefore, there will be points on the warp bubble wall where $\nabla^2 \phi < 0$. At these points, one has $R_{00}<0$, and the strong energy condition is violated (as it we know it must be -- see \cite{N02a}); in the corresponding Newtonian limit, this translates as the mass density generating the gravitational field being negative.
% 
%
%%%%%%%%%%%%%%%%%%%%%%%%%%%%%%%%%%%%%%% Subsection 5.2 %%%%%%%%%%%%%%%%%%%%%%%%%%%%
%
\subsection{Schwarzschild solution}
It is well-known that the Schwarzschild metric can be put in the so-called {\em Painlev\'e-Gullstrand} form
\[
g = - dt \otimes dt + \delta_{ij} \left(dx^i - \frac{(2M)^\frac12 x^i}{r^\frac32} dt\right) \otimes \left(dx^j - \frac{(2M)^\frac12 x^j}{r^\frac32} dt\right)
\]
(see \cite{V04}). Since the vector field ${\bf v}$ with components
\[
v^i = \frac{(2M)^\frac12 x^i}{r^\frac32}
\]
is clearly a gradient, we immediately see that this corresponds to a Newtonian spacetime with potential
\[
\phi = -\frac12 v_i v^i = - \frac{M}r,
\]
which is the Newtonian potential of a point mass $M$, as one would expect. 

Notice that the horizon corresponds to $\|{\bf v}\|=1$; this can be seen as a consequence of the fact that the condition for a timelike curve (using the time coordinate $t$ as the parameter) is $\|\frac{d{\bf x}}{dt}-{\bf v}\|<1$, and explains why the Newtonian formula for the Schwarzschild radius coincides with its General Relativity counterpart.
%
% 
%%%%%%%%%%%%%%%%%%%%%%%%%%%%%%%%%%%%%%% Subsection 5.3 %%%%%%%%%%%%%%%%%%%%%%%%%%%%
%
\subsection{Spatially flat Friedmann-Robertson-Walker models}
Another simple but instructive example is to consider the Newtonian cosmological model generated by the choice
\[
v^i = u(t) x^i = D_i \left(\frac{u(t) r^2}2\right),
\]
corresponding to the potential
\[
\phi = - \frac{\partial}{\partial t} \left(\frac{u(t) r^2}2\right) - \frac12 v_i v^i = - \left(\dot{u} + u^2\right) \frac{r^2}2.
\]
We have $H\phi = -(\dot{u} + u^2) I$ and $K = u I$; consequently,
\[
\left( R_{\mu \nu} \right) =
\left(
\begin{matrix}
-(3\dot{u} + 3u^2) & 0 \\
0 & (\dot{u} + 3u^2) I
\end{matrix}
\right)
\Rightarrow
\left( G_{\mu \nu} \right) =
\left(
\begin{matrix}
3u^2 & 0\\
0 & -(2\dot{u} + 3u^2) I
\end{matrix}
\right)
\]
(see also (\ref{example})). Einstein's equation (\ref{Einstein}) for a dust then yields
\[
\begin{cases}
3u^2 - \Lambda = 8\pi\rho \\
2\dot{u} + 3u^2 - \Lambda = 0
\end{cases}
\]
One can solve the second equation for $u(t)$ and then obtain $\rho(t)$ from the first. These have the following nontrivial physical solutions (the last three with time coordinate chosen so that the big bang occurs at $t=0$):

\begin{enumerate}
\item
{\bf De Sitter universe ($\Lambda > 0$):}
\[
u=\sqrt{\frac{\Lambda}3}, \quad \rho = 0, \quad \phi = -\frac{\Lambda}3\frac{r^2}2.
\]
\item
{\bf Spatially flat Friedmann-Robertson-Walker model with $\Lambda = 0$:}
\[
u = \frac{2}{3t},  \quad \rho = \frac1{6\pi t^2}, \quad \phi = \frac2{9 t^2}\frac{r^2}2.
\]

\item
{\bf Spatially flat Friedmann-Robertson-Walker model with $\Lambda > 0$:}
\[
u = \sqrt{\frac{\Lambda}3} \cotanh\left( \frac12 \sqrt{3\Lambda} t \right),  \quad \rho = \frac{\Lambda}{8\pi \sinh^2\left( \frac12 \sqrt{3\Lambda} t \right)}, \quad \phi = \left( \frac{\Lambda}{6 \sinh^2\left( \frac12 \sqrt{3\Lambda} t \right)} - \frac{\Lambda}3 \right)\frac{r^2}2.
\]

\item
{\bf Spatially flat Friedmann-Robertson-Walker model with $\Lambda < 0$:}
\[
u = \sqrt{\frac{|\Lambda|}3} \cotan\left( \frac12 \sqrt{3|\Lambda|} t \right),  \quad \rho = \frac{|\Lambda|}{8\pi \sin^2\left( \frac12 \sqrt{3|\Lambda|} t \right)}, \quad \phi = \left( \frac{|\Lambda|}{6 \sin^2\left( \frac12 \sqrt{3|\Lambda|} t \right)} + \frac{|\Lambda|}3 \right)\frac{r^2}2.
\]
\end{enumerate}

These are good examples of how Newtonian limits (described by the time-dependent potentials above) can give insight int the nature of the solutions. See \cite{N99} for more examples of Newtonian cosmological solutions of Einstein's field equations.
%
%
%%%%%%%%%%%%%%%%%%%%%%%%%%%%%%%%%%%%% Bibliography &&&&&&&&&&&&&&&&&&&&&&&&&&&&&
%
%

%
%
%%%%%%%%%%%%%%%%%%%%%%%%%%%%%%%%%%%%%%%%%%%%%%%%%%%%%%%%%%%%%%%%%%%%%%%%%%%%
%
%

\begin{thebibliography}{NST99}

\bibitem[Alc94]{A94}
M.~Alcubierre, \emph{The warp drive: hyper-fast travel within general
  relativity}, Class. Quant. Grav. \textbf{11} (1994), L73--L77.

\bibitem[Ehl91]{E91}
J.~Ehlers, \emph{The Newtonian limit of general relativity}, Classical
  mechanics and relativity: relationship and consistency (G.~Ferrarese, ed.),
  Bibliopolis, 1991.

\bibitem[Ehl97]{E97}
\bysame, \emph{Examples of Newtonian limits of relativistic spacetimes}, Class.
  Quant. Grav. \textbf{14} (1997), A119--A126.

\bibitem[LL97]{LL97}
L.~Landau and E.~Lifshitz, \emph{The classical theory of fields},
  Butterworth-Heinemann, 1997.

\bibitem[LV04]{LV04}
F.~Lobo and M.~Visser, \emph{Fundamental limitations on "warp drive"
  spacetimes}, Class. Quant. Grav. \textbf{21} (2004), 5871--5892.

\bibitem[Nat02]{N02a}
J.~Nat\'ario, \emph{Warp drive with zero expansion}, Class. Quant. Grav.
  \textbf{19} (2002), 1157--1165.

\bibitem[NST99]{N99}
P.~Nurowski, E.~Sch\"ucking, and A.~Trautman, \emph{Relativistic gravitational
  fields with close Newtonian analogs}, On Einstein's Path -- Essays in honor
  of Engelbert Sch\"ucking (A.~Harvey, ed.), Springer, 1999.

\bibitem[Tra66]{T66}
A.~Trautman, \emph{Comparison of Newtonian and relativistic theories of
  space-time}, Perspectives in geometry and relativity (B.~Hoffmann, ed.),
  Indiana University Press, 1966.

\bibitem[Vis04]{V04}
Matt Visser, \emph{Heuristic approach to the Schwarzschild geometry}, available
  at gr-qc/0309072 (2004).

\bibitem[Wal84]{W84}
R.~Wald, \emph{General relativity}, University of Chicago Press, 1984.

\end{thebibliography}
\end{document}